\documentclass[fleqn,usenatbib]{mnras}
\usepackage{newtxtext,newtxmath}

\usepackage[T1]{fontenc}

\DeclareRobustCommand{\VAN}[3]{#2}
\let\VANthebibliography\thebibliography
\def\thebibliography{\DeclareRobustCommand{\VAN}[3]{##3}\VANthebibliography}

\usepackage{graphicx}	
\usepackage{amsmath}	
\usepackage{booktabs}
\usepackage{adjustbox}
\usepackage{subcaption}
\usepackage{float}
\usepackage[normalem]{ulem}
\newcommand{\mgii}{Mg~{\sc ii}~}
\newcommand{\civ} {C~{\sc iv}~}

\title[Probing the CGM via \mgii Absorption Coincidence]{Investigating the Circumgalactic Medium through \mgii absorption coincidence}

\author[]{
Paryag Sharma,$^{1}$\thanks{panditparyag@gmail.com}
Raghunathan Srianand,$^{2}$
Hum Chand,$^{1}$
Labanya Kumar Guha$^{3}$
\\
$^{1}$Department of Physics and Astronomical Science, Central University of Himachal Pradesh, Dharamshala, 176215, India\\
$^{2}$Inter-University Centre for Astronomy and Astrophysics, Post Bag 4, Ganeshkhind, Pune 411 007, India\\
$^{3}$Indian Institute of Astrophysics, II Block, Koramangala, Bengaluru-560 034, India
}

\date{Accepted XXX. Received YYY; in original form ZZZ}

\pubyear{\the\year{}}

\begin{document}
\label{firstpage}
\pagerange{\pageref{firstpage}--\pageref{lastpage}}
\maketitle

\begin{abstract}
We present a statistical measurement of the transverse coherence of
Mg\,\textsc{ii} $\lambda\lambda2796,2803$ absorption using a large
sample of 9204 absorber-centric quasar sightline pairs from the Sloan Digital Sky Survey.  
We quantify the probability that an Mg\,\textsc{ii} absorber detected along one
sightline is also present along a nearby sightline, and measure how this
coincidence probability varies with projected separation from $\sim$50\,kpc to
$\sim$1\,Mpc.  
The resulting coincidence curve exhibits a clear two–regime structure: the
coincidence probability rises steeply to $\sim$5--8\% at separations below
$\sim$100\,kpc, but declines rapidly beyond this scale and settles into a
low plateau of $\sim$1--2\% out to $\sim$1\,Mpc.  
A simple geometrical single-halo model reproduces the enhanced probability at
$\lesssim$100\,kpc, while the large-scale plateau is well explained by the
expected contribution from galaxy clustering, confirmed using both photometric
galaxy counts and the two-point correlation function.  
A complementary stacking analysis reveals a significant excess in Mg\,\textsc{ii}
equivalent width in paired sightlines lacking individual detections, implying a
coherence scale of $\sim$100--200\,kpc for the cool, metal-enriched CGM.  
Together, these results identify the transition from a halo-dominated coherence
regime at small separations to a clustering-dominated regime at large scales,
bridging the gap between small-scale lensing constraints and
megaparsec-scale absorber clustering studies.
\end{abstract}




\section{Introduction}
Since the earliest quasar spectra were recorded, intervening metal absorption
lines have revealed the presence of diffuse, ionized and metal enriched gas along the line of
sight \citep{Sandage1965ApJ...141.1560S,Gunn1965ApJ...142.1633G,
Burbidge1966ApJ...144..447B,Burbidge1967ARA&A...5..399B}.  
Subsequent surveys established that many of these absorbers arise from foreground
galaxies intersecting quasar sightlines at small impact parameters\footnote{Impact parameter is the projected physical separation between the center of the galaxy and quasar sight line.} 
\citep[denoted by $b$;][] {Bergeron1991A&A...243..344B, Steidel1995}.
Among these systems, absorption by the Mg\,\textsc{ii} $\lambda\lambda2796,2803$
doublet has become one of the most widely used tracers of cool,
metal-enriched gas associated with the extended circumgalactic medium (CGM) of low-$z$ (i.e., $z<1.0$) galaxies.  In particular, it traces the low ionization (i.e., neutral or singly ionized), low temperature (i.e., $T\sim10^{4}$\,K)  gas with typically high H~{\sc i} column density (i.e., log $N$(H~{\sc i})$\ge$17),
%
%
%
and serves as a key diagnostic of the physical conditions in the
circumgalactic medium (CGM)
\citep[e.g.,][]{Srianand1994}.
The CGM plays a central role in galaxy evolution by mediating the inflow,
outflow, cooling, and recycling of baryons, and Mg\,\textsc{ii} absorbers
provide an effective means to probe these processes over cosmic time \citep{Tumlinson2017ARA&A..55..389T, Fumagalli2024arXiv240900174F, Churchill2025a}.

Absorption-line spectroscopy of background quasars offers a particularly
powerful method for detecting faint CGM gas, as its sensitivity depends primarily
on the achievable signal-to-noise ratio rather than the luminosity of the
foreground galaxy.
However, a single sightline provides a measurement along a pencil-beam through a
complex and anisotropic medium, making it difficult to determine the spatial
extent, morphology, and coherence scale of Mg\,\textsc{ii} absorbing structures.
To overcome this limitation, quasar pairs and multiply imaged gravitationally lensed quasars can be used to
sample multiple closely spaced sightlines through the same foreground structures,
enabling direct measurements of absorber coincidence and transverse variations. \citep[e.g., see][]{Jalan2019ApJ...884..151J,Sharma2025MNRAS.541..601S}.


Using one of the largest samples of quasar pairs available at that time,
\citet{Tytler2009MNRAS.392.1539T} showed that the probability of finding a
coincident metal absorption system declines steeply with increasing transverse
separation—from $\sim20$--$50\%$ at $\lesssim200$\,kpc to $\lesssim1\%$ at
$\sim1$--$2$\,Mpc—consistent with expectations from normal galaxy clustering. 
Similarly, \citet{Rubin2015} have studied coincidence of optically thick absorbers around 40 DLAs at $1.6\le z \le 3.6$ at projected separations $\le$ 300 kpc using closely spaced quasar pairs. 
On the Mpc scale, cross-correlation analysis of \civ\ absorption along widely spaced quasar sightlines are used to constrain effective absorber bias \citep[e.g., see][]{Gontcho2018MNRAS.480..610G}.

%
%

At much smaller scales, lensed quasars and very close quasar pairs allow tests
of the internal coherence of Mg\,\textsc{ii} absorbing structures within
individual haloes. 
Analyses of multi-image lenses show
that strong Mg\,\textsc{ii} and damped Ly$\alpha$ absorbers can vary
substantially between sightlines separated by a few to a few tens of
kpc \citep[e.g.,][]{Ellison2004, Rogerson2012MNRAS.421..971R,Zahedy2016MNRAS.458.2423Z, Rubin2018ApJ...859..146R,Okoshi2021AJ....162..175O, Augustin2021MNRAS.505.6195A}, with
observed equivalent-width differences implying inhomogeneous structure of the absorbing gas
on $\sim$6–12\,kpc scales.
\citet{Rogerson2012MNRAS.421..971R} compared Mg\,\textsc{ii} equivalent widths
across multiple images using Monte Carlo realizations of the halo
model \citep[as in][]{Tinker2008} and found that the observed sightline-to-sightline scatter requires a very
small characteristic coherence length,
$\ell_{c}\!\simeq\!0.5\,h^{-1}\,\mathrm{kpc}$, along with a moderate
covering fraction $f_{c}\!\simeq\!0.6$.  
Large coherence scales ($\ell_{c}\!\gtrsim\!20$\,kpc) were strongly disfavored,
indicating that Mg\,\textsc{ii} absorption can fluctuate substantially even over
sub-kpc separations.
Spatially resolved velocity maps constructed from
multi-image spectroscopy further demonstrate coherent kinematic structures
(e.g., rotating or accreting streams, collimated outflows) on $\sim$5–10\,kpc
scales \citep{Chen2014MNRAS.438.1435C}.  More recent tomographic MUSE studies of
strongly lensed quasars find that the fractional difference in Mg\,\textsc{ii}
equivalent width generally increases with physical separation and suggest a
typical coherence scale of order $\sim$10\,kpc for low-ionization gas, while
high-ionization phases (e.g., C\,\textsc{iv}) appear more coherent at the same
separations \citep{Dutta2024MNRAS.528.1895D}.

Recent work using gravitational-arc tomography provides a complementary, two-dimensional perspective on absorber structure. Spatially resolved spectroscopy of giant arcs has revealed a clumpy, anisotropic \mgii-bearing medium, with absorption strength declining with projected distance and showing strong azimuthal dependence indicative of disc/flow geometry and patchy covering fractions on kpc scales \citep{Lopez2018Natur.554..493L,Afruni2023A&A...680A.112A, Shaban2025}. In particular, \citet{Afruni2023A&A...680A.112A} used VLT/MUSE observations of gravitational arcs to measure the spatial coherence of Mg\,\textsc{ii} absorption across dozens of closely spaced sightlines, finding characteristic coherence scales of order $\sim1$--$8$~kpc. These results reinforce the picture from lensed-quasar studies that low-ionization gas is inhomogeneous on kiloparsec (and sub-kpc) scales and demonstrate the power of extended background sources for mapping absorber morphology.

Together, these works indicate
that Mg\,\textsc{ii} absorption exhibits small-scale clumpiness within halos as
well as larger-scale correlations driven by galaxy clustering, motivating a
systematic study that quantifies coincidence probabilities as a function of
impact parameter, absorber strength and redshift.
%
%
We also would like to understand the
absorber--absorber coincidences on hundred-kpc to
Mpc scales in terms of the radial distribution of gas in the CGM and immediate environments of individual
galaxies at small scales and galaxy-galaxy clustering at large scales \citep[see for example,][]{Tinker2008}.

Modern wide-field spectroscopic surveys such as SDSS \citep{Lyke} provide large samples of quasars spanning the separations necessary to bridge the
gap between the sub-kpc lensed-quasar regime and the Mpc-scale clustering
regime.  
This opens the possibility of performing systematic measurements of
Mg\,\textsc{ii} absorber coincidence as a function of sightline separations, enabling
new constraints on the coherence, covering fraction, and spatial distribution of
cool CGM gas.  
In this work, we exploit a sample of projected quasar pairs with
well-characterized sensitivity and equivalent-width limits to measure the coincident Mg\,\textsc{ii} absorption across tens to hundreds of
kiloparsecs.  
By comparing our results with expectations from galaxy clustering and single
halo-based models, we aim to constrain the physical scales over which 
Mg\,\textsc{ii} absorbing gas remains coherent and to characterize at what scale the
coincidence probability transitions from the single-halo regime to the large-scale galaxy-clustering regime.

This paper is organized as follows: Section~\ref{sec:sample_selection} describes our sample and absorption line measurements. Section~\ref{sec:results} presents our analysis and results, while Section~\ref{sec:discussion} presents the discussion and conclusions of our findings.
Throughout this work, we assume a flat \(\Lambda\)CDM cosmology with \(\Omega_m = 0.3\), \(\Omega_\Lambda = 0.7\), and \(h_0 = 0.7\).


\section{Sample and ABSORPTION LINE MEASUREMENTS}
\label{sec:sample_selection}

\begin{figure*}
    \centering
    \includegraphics[width=\linewidth]{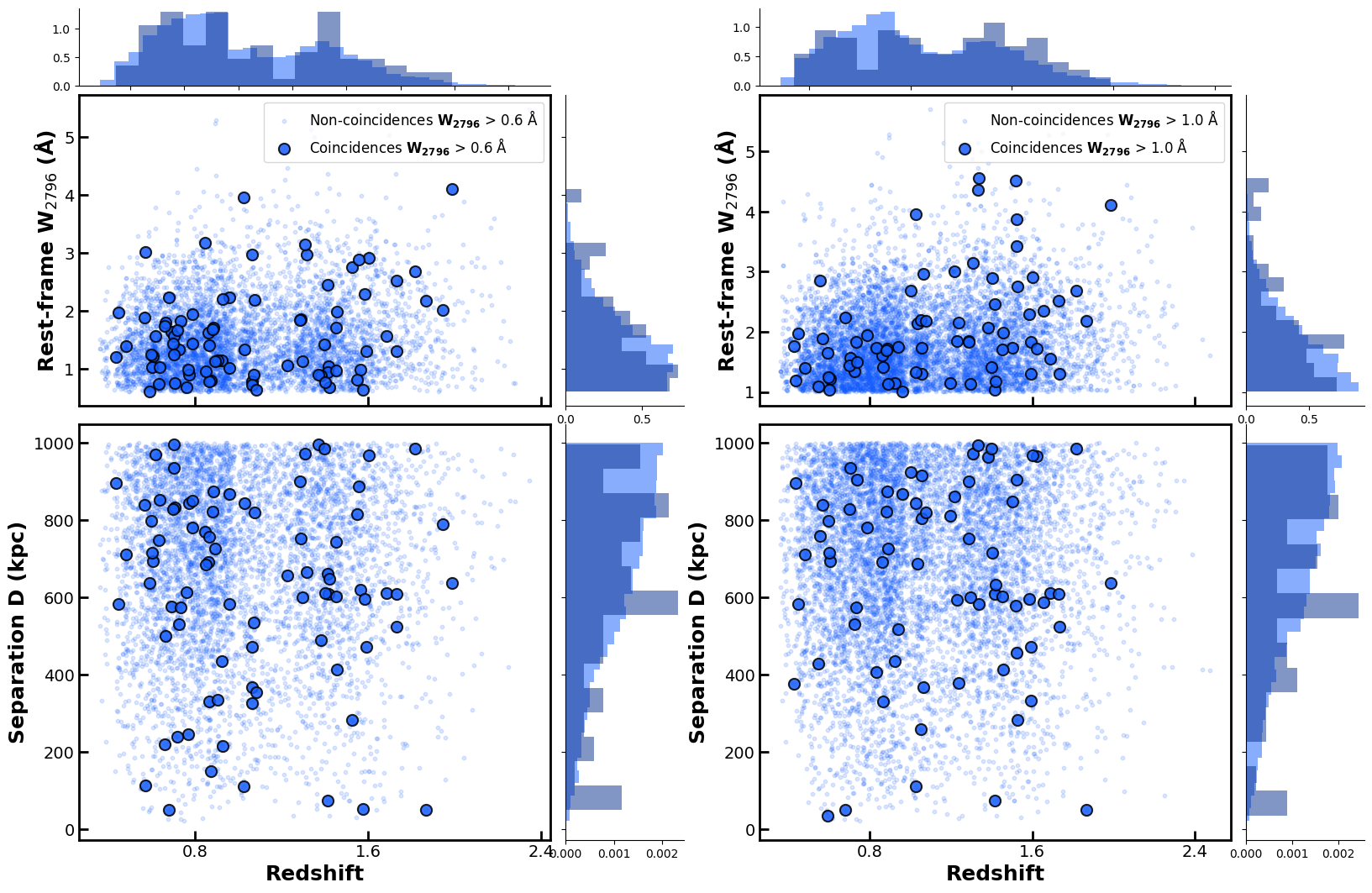}
    \caption{Distribution of pairs for two equivalent
width thresholds ($W_{2796} > 0.6$~\AA\ and $W_{2796} > 1.0$~\AA).
Top panels show the rest-frame equivalent width of detected absorbers,
$W_{2796}$, as a function of absorber redshift, while bottom panels display
the projected separation between the quasar sightlines, $D$, as a function of
absorber redshift. The equivalent width shown corresponds to the absorber
detected along the reference sightline used to initiate the
coincidence search in the companion spectrum. Coincident absorbers are highlighted with larger,
outlined symbols, whereas non-coincident systems are shown with smaller,
fainter points. The marginal histograms along the top and right of each panel
show the normalized distributions of absorber redshift and equivalent width
(top panels) or projected separation (bottom panels), respectively, allowing
a direct visual comparison between the coincident and non-coincident samples
independent of sample size.
}
    \label{fig: ewvsz}
\end{figure*}

We constructed our sample from the SDSS-DR16Q quasar catalog \citep{Lyke}, which contains 750,414 quasars. The initial list of  \mgii absorbers were gathered from the SDSS DR16 \mgii absorber catalog from \citet{Anand2021MNRAS.504...65A}, supplemented with absorbers from \citet{Zhu2013ApJ...770..130Z} to improve completeness. 
Some absorption features that appear closely spaced in wavelength do not correspond to distinct physical systems. Such duplicates arise both from the combination of two independently generated absorber catalogs and from repeated detections of the same system by the automated line–finding pipelines used within each catalog. To address this, we identified and removed duplicate entries, retaining only unique absorption systems detected at a significance of at least $3\sigma$.


As a starting point we use these absorber catalogs for the detection of \mgii systems along individual quasar sightlines.
Each detected \mgii absorber along a sightline is treated as an independent single-sightline case; consequently, a given sightline may contribute multiple such cases if it hosts multiple absorbers. Throughout this work, we refer to these independent absorber-centric quasar sightline cases as “pairs” such that each case is defined by an absorber and its corresponding companion sightline either with or without absorber. We then constructed a pair sample by requiring that \mgii absorber is detected at least in one sightline, with a projected separation (D) of $<1$~Mpc at the absorber redshift. This selection resulted in 38,154 pairs. For each pair, we further imposed two conditions: (1) $z_{\mathrm{abs}} < z_{\mathrm{qso}}$ is satisfied for quasars in both sightlines with a velocity offset greater than $3000~\mathrm{km~s^{-1}}$, and (2) the observed wavelength of the \mgii doublet lies outside the Lyman-$\alpha$ forest region of the companion quasar spectrum. Applying these criteria yielded a final sample of 25,764 pairs.   

Automated absorber catalogs are often incomplete and are therefore commonly supplemented by visual inspection. However, given the large size of our sample, visual inspection is impractical and may introduce subjective biases. We therefore adopt a more robust and fully automated approach based on template matching to confirm the catalog based detection of \mgii absorber in all 25,764 pairs towards both the sightlines. Template matching is a standard technique in which a model absorption
profile is cross-correlated with the observed spectrum to enhance the
detectability of weak or noisy features \citep[e.g.,][]{Zhu2013ApJ...770..130Z}. For this we perform cross-correlation within $\pm1500~\mathrm{km\,s^{-1}}$ around the absorber redshift towards both the sightlines in a pair. A detection was retained only if the maximum correlation peak within the search window exceeded the local mean by at least three times the standard deviation. The corresponding peak position provided an initial estimate of the absorber redshift, which was subsequently refined through Gaussian profile fitting. As an additional consistency check, we verified the doublet nature of \ion{Mg}{II} absorption using an autocorrelation test, requiring significant peaks at the expected rest wavelength separation of $\sim7.18$~\AA. In case of coincidence, the absorbers may be independently cataloged along both sightlines of a pair within $\pm1500~\mathrm{kms^{-1}}$ which can lead to counting the same pair twice. In such cases, we retain only a single entry to avoid duplication. This procedure yielded 19740 pairs.
The equivalent width ($W$) was measured by integrating the observed normalized flux over a $\pm3\sigma$ window centered on the fitted line centroid, where $\sigma$ corresponds to the standard deviation of the best-fit Gaussian profile to the absorption line. To minimize false positives, we required the \ion{Mg}{II} doublet ratio to satisfy $0.7 < W_{2796}/W_{2803} < 2.6$, and the rest-frame full width at half maximum (FWHM) of the line profile to be $<5$~\AA. These criteria lead to 18618 pairs.

For each pixel, the noise equivalent width detection limit was computed using a weighted sum of the line spread function (LSF), modeled as a Gaussian and integrated over a 10-pixel window centered on the pixel. To define coincidences for a given equivalent-width threshold, both sightlines were required to be sensitive to that threshold at more $3\sigma$ level, and each absorber had to exceed the threshold with at least $3\sigma$ significance. For non-coincident cases, both sightlines met the $3\sigma$ sensitivity criterion, but only one exhibited an absorber above the threshold at $\geq3\sigma$ significance.
For equivalent-width thresholds of $1.0$ and $0.6$~\AA, we identify $71$ and $82$ coincident pairs out of total samples of $7719$ and $5512$ quasar pairs, respectively with 4027 pairs common among them. 
All these details are summarized in Table~\ref{tab:samdet}, and the full sample of $9204$ unique pairs is listed in Table~\ref{tab:mgii_coincidence_sample}.

Figure~\ref{fig: ewvsz} shows the distribution of the rest-frame Mg\,\textsc{ii}
equivalent width ($W_{2796}$) and projected quasar-pair separation ($D$) as a
function of absorber redshift for two equivalent-width thresholds,
$W_{2796} > 0.6$~\AA\ (left panels) and $W_{2796} > 1.0$~\AA\ (right panels).
Coincident absorbers are highlighted with filled symbols, while
non-coincident systems are shown with lighter points.
The marginal histograms shown along the top of each panel represent the
normalized redshift distributions of the coincident and non-coincident
samples, while the histograms along the right-hand side show the normalized
distributions of rest-frame equivalent width (top panels) or projected
separation (bottom panels). In all cases, the histograms are normalized to
unit area to facilitate a direct visual comparison between the two samples,
independent of their differing sample sizes.
To quantitatively assess whether the coincident and non-coincident systems are
drawn from the same underlying distributions, we performed Kolmogorov--Smirnov
(KS) tests on both the projected separation and absorber redshift
distributions. For the $W_{2796} > 0.6$~\AA\ sample, the KS test yields
$p = 0.08$ for the separation distribution and $p = 0.41$ for the redshift
distribution. Similarly, for the $W_{2796} > 1.0$~\AA\ sample, we obtain
$p = 0.16$ for separation and $p = 0.09$ for redshift. In both cases, the KS
tests do not reject the null hypothesis that the coincident and non-coincident
samples are drawn from the same parent distributions.
This result indicates that the global redshift and separation distributions of
coincident absorbers are statistically consistent with those of the
non-coincident population. However, this does not preclude a relative excess of
coincident absorbers at small projected separations. The KS test is sensitive
to differences across the full distribution and may not capture localized
enhancements at small $D$, which are more directly probed by the binned
coincidence probabilities and the halo-scale modeling presented in subsequent
sections.

\begin{figure}
    \centering
    \includegraphics[width=1\linewidth]{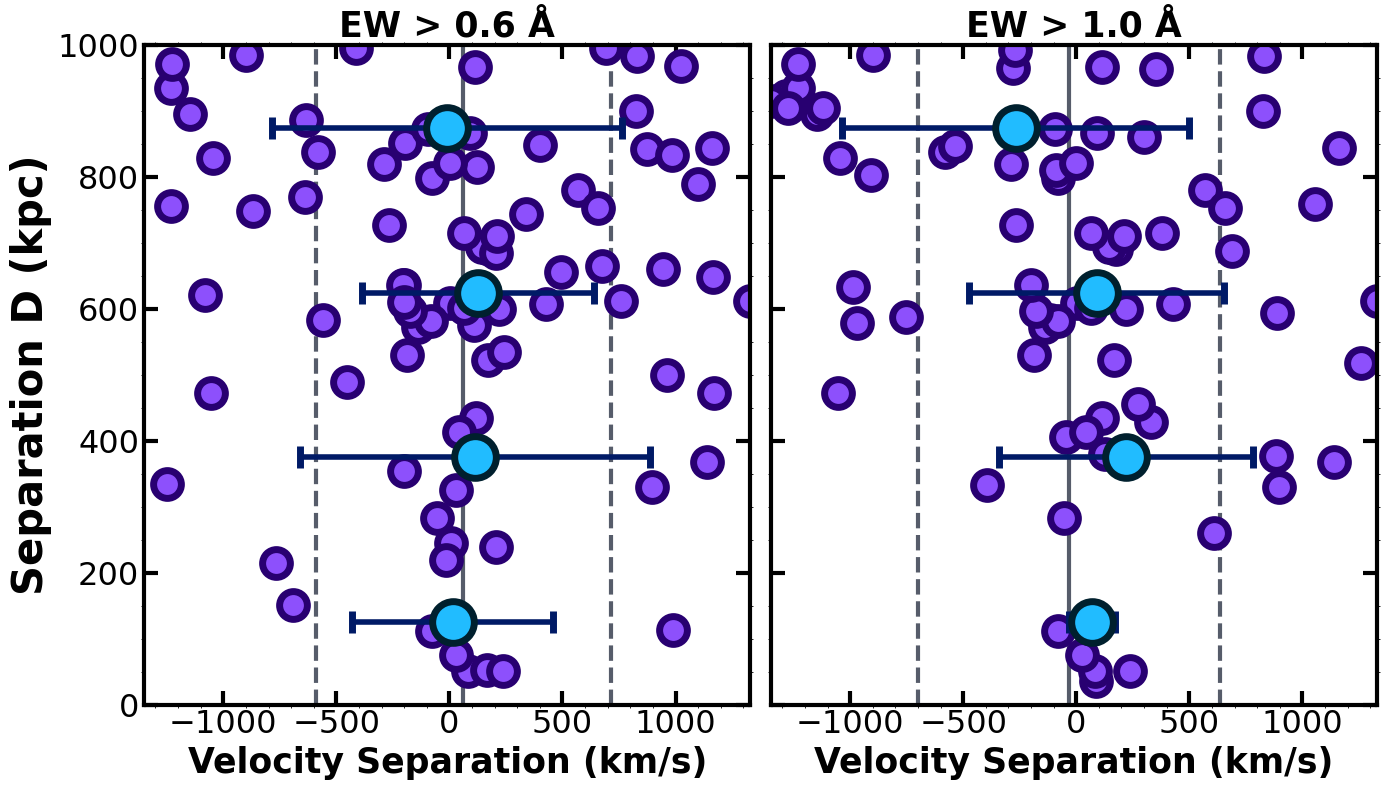}
    \caption{Scatter plots of line-of-sight velocity separation as a function of projected
separation for coincident Mg\,\textsc{ii} absorbers. The two panels show results
for equivalent-width sensitivity thresholds of $W_{2796} > 0.6$~\AA\ (left) and
$W_{2796} > 1.0$~\AA\ (right). Individual points represent measured velocity
offsets for coincident systems, while the overplotted symbols with horizontal
error bars indicate the mean velocity separation in bins (250 kpc) of projected
separation D, with the error bars denoting the
corresponding dispersion within each bin. The solid vertical line marks the mean velocity separation of the full sample, while the dashed lines indicate the $\pm1\sigma$ dispersion from a Gaussian fit to the velocity distribution.
}
    \label{fig:velocity-offset}
\end{figure}

Figure~\ref{fig:velocity-offset} shows the line-of-sight velocity separations of
coincident Mg\,\textsc{ii} absorbers that satisfy the $W_{2796} > 0.6$~\AA\
(left panel) and $W_{2796} > 1.0$~\AA\ (right panel) thresholds as a function of
projected separation between the quasar sightlines. Individual points
represent the measured velocity offsets, while the larger symbols with
horizontal error bars indicate the mean velocity separation in bins of
projected separation, with the error bars denoting the corresponding
dispersion. Two things are evident from this figure. Firstly, at a given projected
separation, the velocity off-set has a small scatter for the strong absorbers at lower projected separation (D). Secondly the spread in velocity off-set increases with increasing projected separation for a given equivalent width cutoff. 

\begin{figure}
    \centering
    \includegraphics[width=\linewidth=]{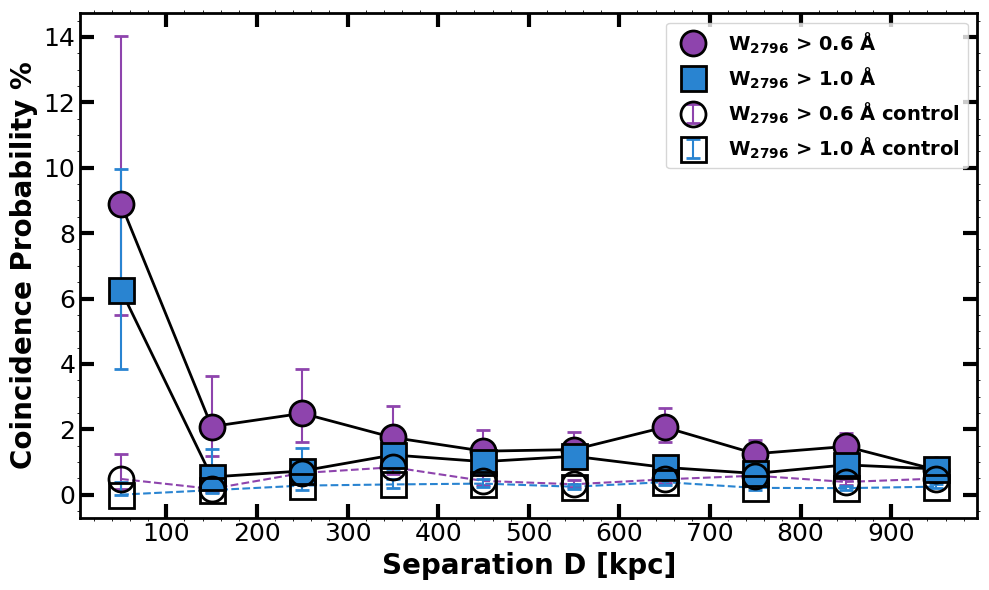}
    \caption{Coincidence probability (\%) as a function of projected pair separation $D$ at the absorber redshift. It is shown for two equivalent-width thresholds, $W_{2796} > 0.6$~\AA\ (violet circles) and $W_{2796} > 1.0$~\AA\ (blue squares). The corresponding control samples, are shown with unfilled symbols. Error bars represent binomial uncertainties corresponding to a 68\% ($1\sigma$) confidence interval.}
    \label{fig:coincidence probability}
\end{figure}

Figure~\ref{fig:coincidence probability} shows the ratio of coincident pairs to total pairs as a function of projected separation (D) at $z_\mathrm{abs}$ for these two equivalent width thresholds. The uncertainties on the coincidence probabilities were estimated using binomial statistics, corresponding to a 68\% ($1\sigma$) confidence level. We constructed a control sample by randomly reassigning quasar partners while preserving the absorber redshift and projected separation. The same analysis was repeated on this control sample, and the resulting coincidence probabilities as a function of separation are shown in Figure~\ref{fig:coincidence probability} for comparison.

\begin{table}
\centering
\caption{Summary of sample selection. Counts are reported in terms of “pairs,” where each pair corresponds to an absorber-defined case consisting of one sightline with a detected absorber and its corresponding companion sightline, which may or may not exhibit absorption. Hence, a given quasar sightline may therefore contribute multiple pairs if it hosts multiple absorbers.}
{\tiny

\begin{tabular}{c l r}
\hline\hline
\textbf{S. No.} & \textbf{Selection Step} & \textbf{Count} \\
\hline
(1)  & SDSS-DR16Q quasars & 750,414 \\
(2)  & The pairs in (1) satisfying $\le 1000$ kpc and an absorber in at least one sightline & 38154 \\
(3)  & The pairs in (2) satisfying  $z_{\rm abs}<z_{\rm qso}$, $\Delta v>3000$ km\,s$^{-1}$, no Ly$\alpha$ overlap & 25764 \\
(4)  & The pairs in (3) satisfying the cross + auto-correlation criteria either in one or both sightlines & 19740 \\
(5)  & The pairs in (4) satisfying FWHM $<5$\,\AA, doublet ratio 0.7--2.6 either in one or both sightlines  & 18618 \\

\multicolumn{3}{c}{Pairs sample with SNR satisfying for threshold of $W^{\rm th}_{2796}>1.0$\,\AA} \\
(6)  & The pairs in (5) with spectra SNR satisfying $W_{2796}>1.0$\,\AA\ in both and & 7719 \\

 & absorption either in one or both sightlines & \\
(7)  & The pairs in (5) with spectra SNR satisfying $W_{2796}>1.0$\,\AA\ in both and absorption  & 71 \\\
 &in both sightlines (Coincidence)& \\
\multicolumn{3}{c}{Pairs sample with SNR satisfying for threshold of $W^{\rm th}_{2796}>0.6$\,\AA} \\
(8)  & The pairs in (5) with spectra SNR satisfying $W_{2796}>0.6$\,\AA\ in both and & 5512 \\
&absorption either in one or both sightlines &\\
(9)  & The pairs in (5) with spectra SNR satisfying $W_{2796}>0.6$\,\AA\ in both and absorption  & 82 \\
 &in both sightlines (Coincidence)& \\

\hline\hline
\end{tabular}
}
\label{tab:samdet}
\end{table}

\begin{table*}
\centering
\caption{
Our final sample consists of $9204$ pairs as summarized in Table \ref{tab:samdet}. Of these, $1485$ pairs are used only for the analysis with an equivalent-width threshold of $\mathrm{EW} \geq 0.6\,\text{\AA}$, $3692$ pairs are used only for the $\mathrm{EW} \geq 1.0\,\text{\AA}$ analysis, and $4027$ pairs are common to both samples as listed below. Out of the total $5512$ pairs used in the $\mathrm{EW} \geq 0.6\,\text{\AA}$ sample, $82$ show coincident absorption. Similarly, among the $7719$ pairs used in the $\mathrm{EW} \geq 1.0\,\text{\AA}$ sample, $71$ exhibit coincidence.
The \textit{Thresh.\ Flag} column denotes the equivalent-width threshold satisfied by each system:
First sample (S$_1$) with  $W_{2796}>0.6$~\AA, and second sample (S$_2$) with $W_{2796}>1.0$~\AA, and S$_1$S$_2$ (both).
The \textit{Coinc.\ Flag} column indicates whether Mg\,\textsc{ii} absorption satisfying the corresponding threshold is detected in both sightlines (1) or not (0).
For non-coincident systems, only the sightline containing detected absorption is listed; columns corresponding to the other sightline without detection are assigned a placeholder value of $-999$.
Only a subset of columns and rows is shown here for illustration; the full table is available online.
}
\label{tab:mgii_coincidence_sample}
\tiny
\begin{tabular}{l c l c c c c c c c c c c c c}
\hline\hline
QSO$_1$ & $Z_{\rm QSO,1}$  & QSO$_2$ & $Z_{\rm QSO,2}$ & $D$ & $z_{\rm abs,1}$ & $W_{2796,1}$ & $\sigma_{W,1}$ & $z_{\rm abs,2}$ & $W_{2796,2}$ & $\sigma_{W,2}$ & EW$_{\rm lim,1}$ & EW$_{\rm lim,2}$ & Thresh. Flag & Coinc. Flag \\
 &  &  &  & (kpc) &  & (\AA) & (\AA) &  & (\AA) & (\AA) & (\AA) & (\AA) &  &  \\
 (1) & (2) & (3) & (4) & (5) & (6) & (7) & (8) & (9) & (10) & (11) & (12) & (13) & (14) & (15) \\
\hline
000014.12$-$030936.5 & 1.9768 & 000016.08$-$030809.8 & 1.9821 & 737.9 & $-999$ & $-999$ & $-999$ & 1.0299 & 1.04 & 0.20 & 0.0655 & 0.1049 & S1S2 & 0 \\
000024.62+204530.4 & 1.013 & 000030.10+204617.9 & 1.2125 & 701.2 & 0.8886 & 1.62 & 0.42 & $-999$ & $-999$ & $-999$ & 0.2140 & 0.1263 & S2 & 0 \\
000036.48+321102.4 & 3.221 & 000041.24+321032.9 & 1.8672 & 535.2 & $-999$ & $-999$ & $-999$ & 0.9721 & 3.05 & 0.20 & 0.2193 & 0.1278 & S2 & 0 \\
$\vdots$ & $\vdots$ & $\vdots$ & $\vdots$ & $\vdots$ & $\vdots$ & $\vdots$ & $\vdots$ & $\vdots$ & $\vdots$ & $\vdots$ & $\vdots$ & $\vdots$ & $\vdots$ & $\vdots$ \\
\hline
\end{tabular}
\end{table*}

\section{Results}
\label{sec:results}

The curves shown in Figure~\ref{fig:coincidence probability} exhibit two distinct regimes: a steep rise in the coincidence probability at $D\lesssim 100$~kpc, followed by a nearly flat plateau at larger separations. At small impact parameters ($D \lesssim 100$~kpc), the observed coincidence probabilities are $0.088^{+0.051}_{-0.033}$ for $W_{2796}>0.6$~\AA\ and $0.062^{+0.037}_{-0.023}$ for $W_{2796}>1.0$~\AA. This excess at small separations is indicative of absorption arising on halo scales, which we model in the following sections as a single-halo contribution.
At larger separations ($D \gtrsim 100$~kpc), the coincidence probability flattens to mean value of $0.0162 \pm 0.0053$ for $W_{2796}>0.6$~\AA\ and $0.0087 \pm 0.0023$ for $W_{2796}>1.0$~\AA, where the quoted uncertainties across radial bins corresponding to a 68\% ($1\sigma$) confidence level using binomial statistics. In this regime, coincident absorption is dominated by correlated but distinct galaxies tracing the same large-scale structure and is therefore modeled as a clustering-driven contribution. For comparison, the corresponding control samples yield significantly lower mean probabilities of $0.0048 \pm 0.0019$ and $0.0026 \pm 0.0008$ for the same equivalent-width thresholds, confirming that the observed plateau reflects genuine clustering rather than chance alignments.
We also examine the coincidence probability of our sample without imposing any equivalent-width threshold ($W_{2796}>0$), retaining only the $3\sigma$ detection requirement. In this case, $D<100$~kpc bin shows a coincidence probability of $0.038^{+0.018}_{-0.012}$, while the average probability at $D>100$~kpc is $0.0088 \pm 0.0025$. This confirms the trend found for the equivalent width limited samples. 
The lower amplitude relative to the equivalent-width–limited samples arises because our sample consists of large number of low SNR spectra where only very strong lines are detectable. These spectra contribute to the denominator without proportionately contribute to the numerator.
As a result, the $W_{2796}>0$ sample should 
only be treated as a consistency check.




\subsection{Modeling Coincident Probability due to single halo}

The rest-frame equivalent width of \mgii\ absorption is known to anti-correlate with impact parameter (b), though with significant scatter \citep{Bergeron1991A&A...243..344B,Chen2010ApJ...714.1521C,Nielsen2013ApJ...776..114N}. This relation is typically modeled as a log-linear function:
\begin{equation}
    \log W_{2796}\ (\text{\AA}) = \alpha_{1} b\ (\text{kpc}) + \beta_{1}
    \label{eq:w-d}
\end{equation}
where $\beta_{1}$ is the intercept at $b = 0$ and $\alpha_{1}$ defines the exponential scale. For the redshift range $0 < z < 1.5$, we adopt $\alpha_{1} = -0.019 \pm 0.002$ and $\beta_{1} = 0.540 \pm 0.028$ from \citet{Guha2024MNRAS.527.5075G}. Since $82.63\%$ of the absorbers in our sample lie within this redshift interval, this is well suited for our analysis. The probability that one will detect Mg~{\sc ii} absorption at a given $b$ can be obtained from the measured covering factor ($f_c$) as a function of $b$.
This can be modeled as:
\begin{equation}
    f_c = A_{100} \left(\frac{b}{100\ \text{kpc}}\right)^{\gamma},
    \label{eq: covering frac}
\end{equation}
Here $A_{100}$ is given as :
\begin{equation}
A_{100} = A \times (1 + z)^{\alpha_{2}} \left( \frac{M_*}{10^{10} M_\odot} \right)^{\beta_{2}}.
\end{equation}
We adopt the best-fit parameters $\gamma$ and $A_{100}$ from \citet{Lan2020ApJ...897...97L}, using their average values derived for star-forming galaxies. The normalization $A_{100}$ corresponds to typical $L^*$ galaxies with a stellar mass of $M = 10^{10.8} M_{\odot}$ at redshift $z = 1$, consistent with the median absorber redshift in our sample.

\begin{figure}
    \centering
    \includegraphics[width=1\linewidth]{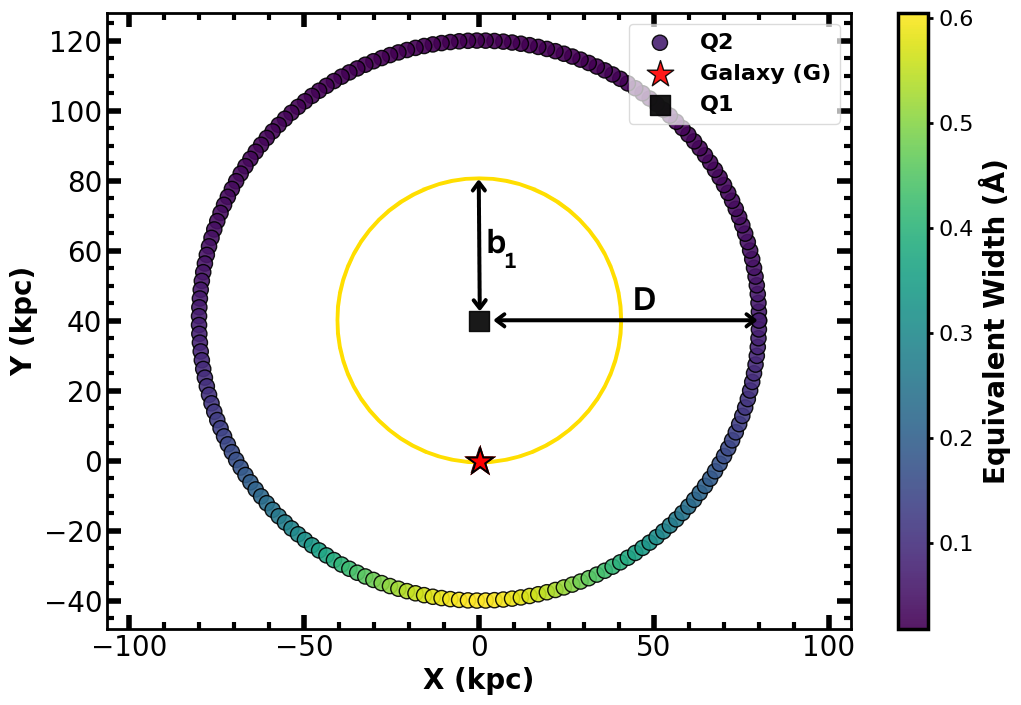}
    \includegraphics[width=0.95\linewidth]{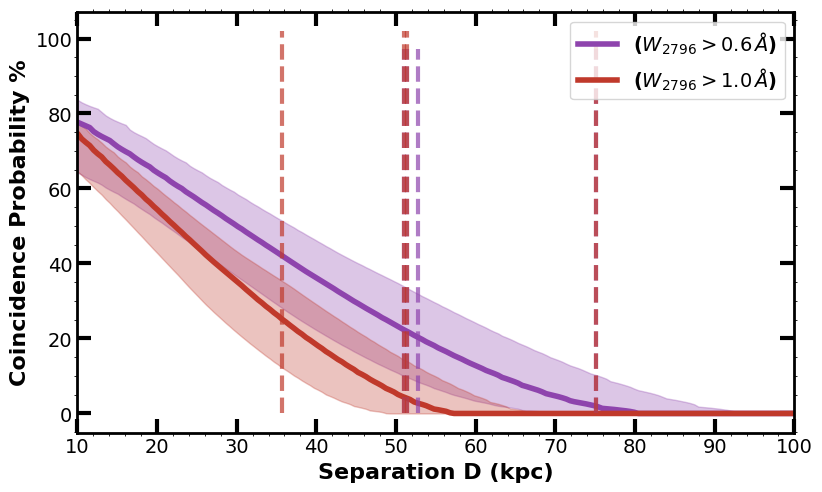}
    \caption{Top panel: Illustrating an example of modeling coincidence probability using a single halo. Suppose the Q1 sightline shows a \mgii\ 2796 absorption with $W_{2796} = 0.6$\,\AA. Using the $W$--$D$ relation, we infer the galaxy lies at an impact parameter of 40.10 kpc. The plot then shows all possible Q2 positions at projected separation (60 kpc) from the quasar Q1, and the corresponding $W_{2796}$ values computed using the same relation. Bottom panel: Variation of the coincidence probability, $P_{\mathrm{coinc}}$, with the projected separation between Q1 and Q2 at the absorber redshift. The vertical lines indicate the projected separations of the observed absorbers at which the coincidence–probability curve was sampled.}
    \label{fig:single_halo}
\end{figure}

Let Q1 and Q2 are the two quasars which are separated by a small projected separation (D) in the sky.
For an \mgii absorber of rest-equivalent width $W_{2796}$ detected in the spectrum of Q1, the
$W$--$b$ relation (Eq.~\ref{eq:w-d}) implies a corresponding galaxy at an impact
parameter $b_{1}$. The galaxy may therefore lie anywhere along a
circle of radius $b_{1}$ centered on Q1, as illustrated in the top
panel of Figure~\ref{fig:single_halo}. For a quasar pair with projected
separation $D$, each possible galaxy position generates a different
galaxy--Q2 impact parameter ($b_{2}$). This is geometrically equivalent to fixing the
galaxy at one location and allowing Q2 to move along a circle of radius $D$
around Q1. Applying the $W$--$b$ relation to each such configuration yields a
predicted equivalent width along Q2. The coincidence probability at separation
$D$ is then given by the fraction of the circle for which the predicted width
exceeds the adopted threshold, multiplied by the covering fraction
(Eq.~\ref{eq: covering frac}).

To incorporate the distribution of observed absorber strengths, we repeat this
procedure over a range of initial equivalent widths $W_{1}$ detected in Q1.
For the $W_{2796} > 0.6$~\AA and $W_{2796} > 1.0$~\AA\ samples, we sample
$W_{1}$ uniformly using 20 equally spaced values over the ranges
$0.6$--$3.0$~\AA and $1.0$--$3.0$~\AA, respectively. Each
$W_{1}$ maps to an impact parameter $b_{1}$ via
Eq.~\ref{eq:w-d}, producing a set of coincidence probabilities for each $D$.
The final coincidence probability curve is obtained by taking the weighted mean of these individual curves, using the exponential equivalent width distribution
\begin{equation}
  p(W_{2796}) = \frac{1}{W_*} \exp\!\left(-\frac{W_{2796}}{W_*}\right),
  \label{eq:ew_prob}  
\end{equation}
as the weighting function, with $W_* = 0.702 \pm 0.017$~\AA~ characterizing the observed distribution of Mg\,\textsc{ii} absorber strengths \citep[see][]{Nestor2005ApJ...628..637N}. This weighted mean represents the expected probability of detecting coincident absorption as a function of projected separation (D) and detection threshold.
Uncertainties are propagated by varying all model parameters within their $1\sigma$ bounds.
The resulting upper and lower bounds define the shaded uncertainty region in the bottom panel of Figure~\ref{fig:single_halo}.

If we sample the modeled coincidence probability curves at the observed projected quasar separations, 
the mean probabilities for the $W_{2796} > 0.6~\text{\AA}$ and $W_{2796} > 1.0~\text{\AA}$ thresholds are 
$0.149^{+0.112}_{-0.087}$ and $0.069^{+0.083}_{-0.045}$, respectively. 
The corresponding observed coincidence probabilities are 
$0.088^{+0.051}_{-0.033}$ and $0.062^{+0.037}_{-0.023}$, respectively which agree to the modeled probabilities within $1 \sigma$ level.
To investigate whether the coincidence probability evolves with redshift, we repeated the analysis at two additional redshifts corresponding to the lower 25th and higher 75th percentiles of our sample, located at $z = 0.73$ and $z = 1.35$, respectively. At $z = 0.7278$, the predicted mean coincidence probabilities are $0.133^{+0.105}_{-0.077}$ for the $W_{2796} > 0.6$~\AA\ threshold and $0.056^{+0.072}_{-0.037}$ for the $W_{2796} > 1.0$~\AA\ threshold. At the redshift of $z = 1.3688$, these values increase to $0.166^{+0.108}_{-0.096}$ and $0.084^{+0.075}_{-0.053}$, respectively. These results reveal a mild but systematic increase in coincidence probability with redshift, and since the covering fraction itself is known to increase with redshift, this naturally drives the observed rise in coincidence probability.

\subsection{Modeling the Coincident Probability due to clustering}
While the single-halo model naturally explains the rapid rise in coincidence probability at separations below $\sim 100$~kpc, the observed plateau at larger separations must arise from a different physical mechanism. On scales of several hundred kiloparsecs to a few megaparsecs, \mgii absorbers are expected to correlate with the surrounding galaxy population through large-scale structure \citep{Bouche2004MNRAS.354L..25B,Lundgren2009ApJ...698..819L,Gauthier2009ApJ...702...50G}.
To estimate the coincidence probability due to galaxy clustering, we use pairs as reference sightlines. First we address this problem using available photometric redshift catalog of galaxies.

\begin{figure}
    \centering
    \includegraphics[width=0.9\linewidth]{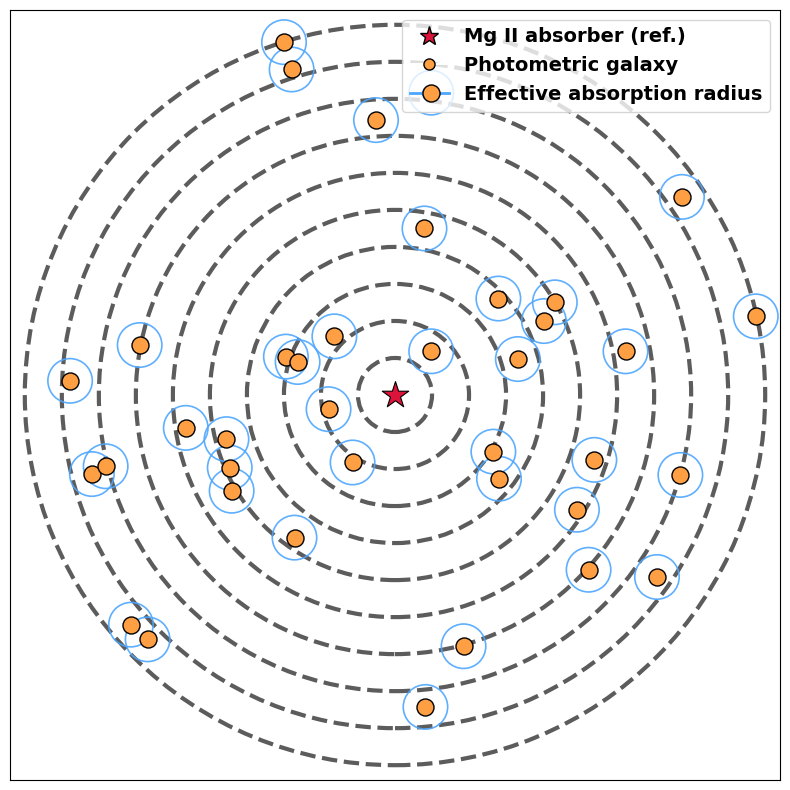}
    \includegraphics[width=\linewidth]{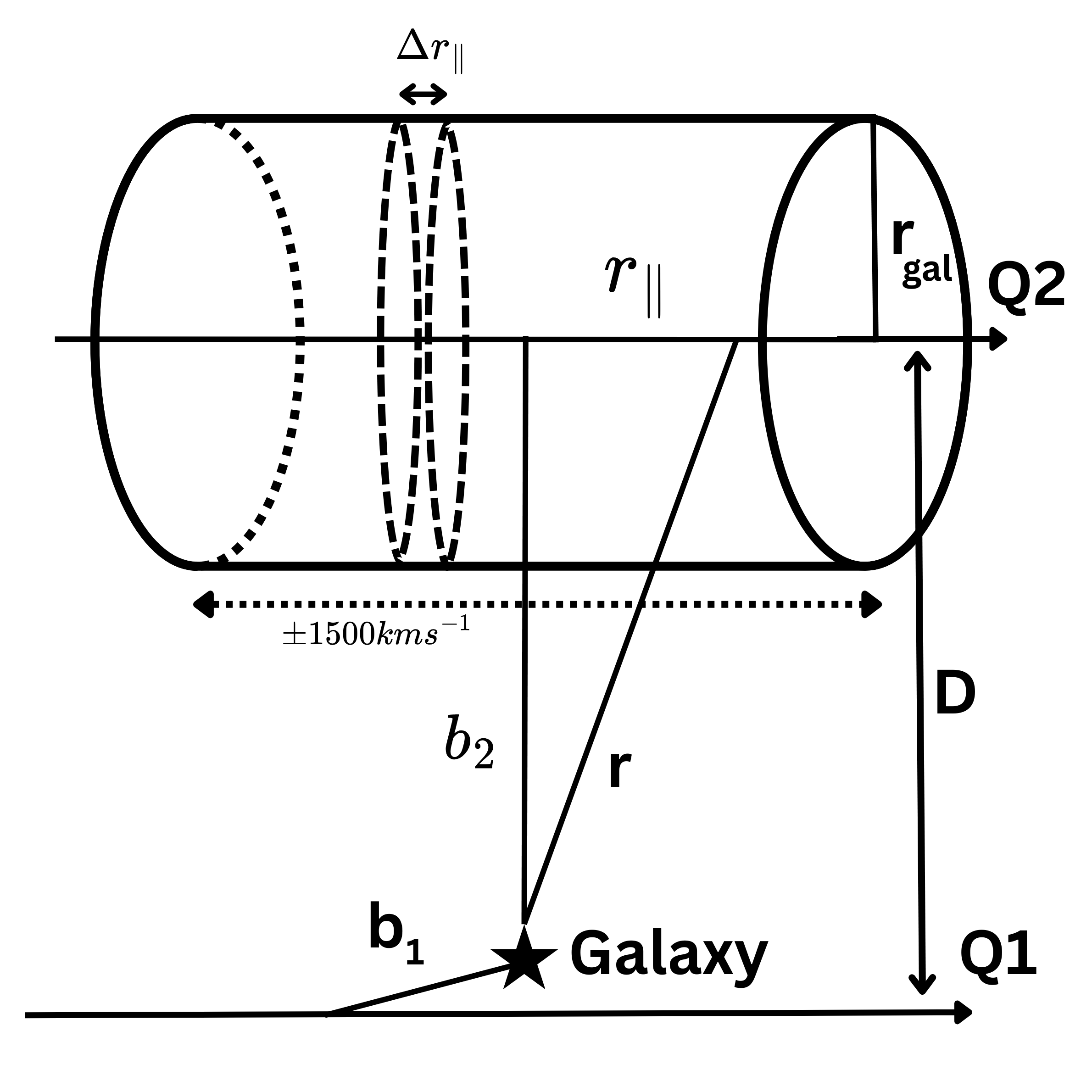}

\caption{Top Panel: Schematic illustration of the projected–shell method used to estimate the coincidence probability arising from galaxy clustering. Concentric annuli (100 kpc width) are drawn around each reference Mg\,II absorber. Photometric galaxies from the Legacy Survey DR10 within the redshift interval $0.4<z_{\rm abs}<0.7$ are shown with their corresponding effective absorption radii (40 kpc and 20 kpc for equivalent-width thresholds of 0.6 Å and 1.0 Å, respectively). Only absorbers within this redshift range and within the DR10 footprint are used in the analysis.
Bottom panel: Schematic illustration of the clustering-based model used to estimate the theoretical coincidence probability.
Given an Mg\,\textsc{ii} detection in the first sightline (Q1) at impact parameter $b_{1}$, a cylindrical volume is constructed around the second sightline (Q2) with radius $r_{\rm gal}(W_{2796})$ and a line-of-sight half-length corresponding to $\pm1500~\mathrm{km\,s^{-1}}$.
The expected number of galaxies within this cylinder is computed by integrating the galaxy two-point correlation function $\xi(r) = (r/r_{0})^{-\gamma}$, and is interpreted as the probability of detecting a coincident absorber as a function of the projected separation $D$.}

\label{fig:clustering_coincidecne_probability_diagrams}
\end{figure}

For each pair, the quasar’s position and absorber redshift define a cylindrical search volume of radius 1000~kpc and line-of-sight depth corresponding to a velocity range of $\pm1500~\mathrm{km\,s^{-1}}$, converted to a redshift interval around the absorber redshift (see the schematic illustration in the top panel of Figure~\ref{fig:clustering_coincidecne_probability_diagrams}).
In constructing these cylinders, we use the RA, Dec, and absorber redshift ($z_{\rm abs}$) from our quasar sample and restrict the analysis to systems with $0.4 < z_{\rm abs} < 0.7$, corresponding to the redshift range where the photometric galaxy catalog from the DESI Legacy Imaging Surveys DR10 remains well characterized in terms of depth and completeness \citep{Dey2019AJ....157..168D}. The survey provides wide-area optical imaging reaching typical $5\sigma$ point-source depths of $g \simeq 24.0$, $r \simeq 23.4$, and $z \simeq 22.5$ AB magnitudes for faint galaxies.
We then count photometric galaxies that fall within these cylindrical volumes. 
Since photometric redshifts carry significant uncertainties, typically 
$\sigma_{\rm NMAD}$ (Normalized Median Absolute Deviation) $\sim 0.01$--$0.02$ 
when compared with spectroscopic redshifts 
\citep{Li2024AJ....168..233L}, each galaxy is treated as a Gaussian in redshift space, with its mean given by the photometric redshift and standard deviation equal to its reported photo-$z$ error. 
For each absorber, we compute the probability that a given photometric galaxy lies within the absorber’s redshift range by integrating this Gaussian across the corresponding $\Delta z$ interval ($\pm1500~\mathrm{kms^{-1}}$). The contribution of each galaxy to the total probability is thus weighted by this overlap fraction.

\begin{figure*}
    \centering
    \includegraphics[width=\linewidth]{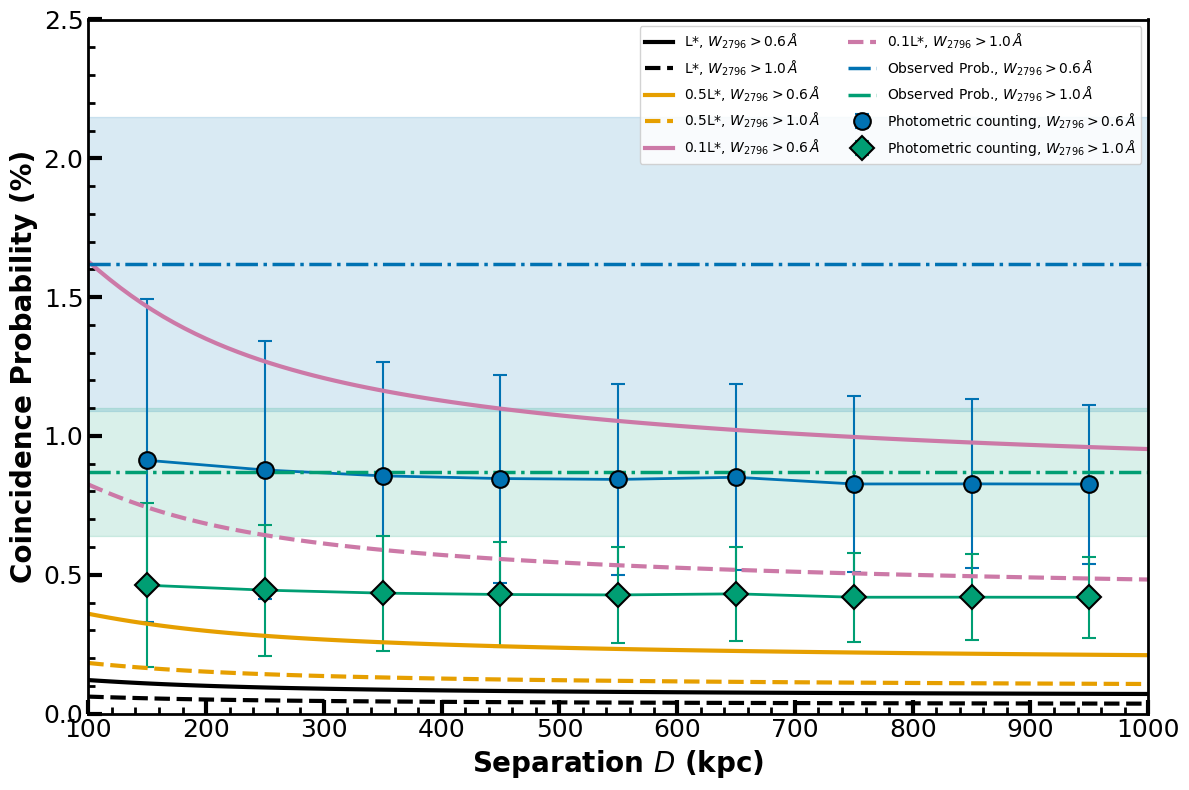}
\caption{
Coincidence probability due to clustering of Mg\,\textsc{ii} absorbers as a function of projected separation $D$ between quasar sightlines.
The horizontal dot--dashed lines and shaded regions indicate the average observed coincidence probabilities and their $1\sigma$ uncertainties for absorbers with $W_{2796} > 0.6$\,\AA\ and $W_{2796} > 1.0$\,\AA~respectively.
Points with error bars show the coincidence probability measured in radial bins from photometric galaxy counting for the same equivalent--width thresholds, corrected for the mean covering fraction.
Solid and dashed curves represent theoretical predictions from the clustering model for $W_{2796} > 0.6$\,\AA\ and $W_{2796} > 1.0$\,\AA, respectively.
For each equivalent--width threshold, the black, orange, and magenta curves correspond to models assuming characteristic host galaxy luminosities of $L^{\ast}$, $0.5L^{\ast}$, and $0.1L^{\ast}$, respectively, where the expected number of galaxies is computed by integrating the galaxy two--point correlation function within the cylindrical search volume.
}
    \label{fig:clustering_curves}
\end{figure*}

To quantify the radial dependence within 1000 kpc, we divide the projected separation into concentric cylindrical shells of thickness 100~kpc. 
Within each shell, we compute the probability-weighted volume fraction occupied by galaxies using
\begin{equation}
p_{\mathrm{sum,vol}} = \sum_i P_i\,(\pi\,r_{\mathrm{gal}}^2\,h),
\end{equation}
where $P_i$ is the overlap probability for the $i$-th galaxy, $r_{\mathrm{gal}}$ is the physical radius associated with the equivalent-width threshold (28.42~kpc for $\mathrm{EW}>1~\text{\AA}$ and 40.10~kpc for $\mathrm{EW}>0.6~\text{\AA}$, derived from the observed $W$--$b$ relation from Equation \ref{eq:w-d}), and $h$ is the line-of-sight height of the absorber cylinder. 
This weighted volume is then normalized by the total shell volume,
\begin{equation}
V_{\mathrm{shell}} = \pi\,h\,(D_{\mathrm{outer}}^2 - D_{\mathrm{inner}}^2),
\end{equation}
yielding the fractional volume within the shell expected to be occupied by galaxies capable of producing absorption. 
Finally, multiplying this fractional volume by the average covering fraction for the corresponding equivalent-width limit gives the probability of detecting an absorber along a background sightline---that is, the coincidence probability due to galaxy clustering.
The resulting coincidence probability profile (Figure~\ref{fig:clustering_curves}) is approximately constant with projected separation, consistent with expectations for galaxy clustering on scales below $\sim 1$~Mpc. However, the predicted probabilities are systematically lower by roughly a factor of two compared to the observed coincidence fractions for both equivalent-width thresholds. This discrepancy likely reflects two limitations of the methodology.

First, the Legacy Survey photometric catalog is limited to relatively luminous galaxies. \cite{Wang2021SCPMA..6489811W} constructed a volume-limited galaxy samples from the DESI Legacy Imaging Surveys (DR8) and report absolute magnitudes in the AB system, showing that at intermediate redshifts ($0.4 \lesssim z \lesssim 0.7$) the samples are complete only down to $M^{0.5}_z - 5\log h \simeq -21$. For $h=0.7$, this corresponds to $M^{0.5}_z \simeq -21.77$. The superscript “0.5” denotes that the magnitudes are $k$-corrected to a reference redshift $z=0.5$. At this redshift, the observed $z$ band ($\lambda_{\rm eff}\approx9100$~\AA) samples rest-frame optical wavelengths ($\lambda_{\rm rest}\approx6000$~\AA), close to the Johnson $V$ band. We therefore interpret $M^{0.5}_z$ as tracing rest-frame $V$-band luminosity to first order, and convert to the $B$ band using a typical rest-frame color $(B-V)\simeq0.6$ for typical galaxies \citep[e.g.,][]{Fukugita1995PASP..107..945F}. Finally, we transform from the AB to the Vega system using the empirical calibration $B_{\rm Vega}=B_{\rm AB}+0.163$ \citep{Frei1996AJ....111..174F}. This corresponds to an approximate rest-frame Johnson $B$-band magnitude of $M_B^{\rm Vega}\simeq-21$. As a result, the Legacy Survey catalog systematically misses fainter galaxies, which are known to contribute significantly to the population of \mgii absorber hosts.
Second, our treatment of photometric redshifts approximating each galaxy as a Gaussian in redshift space and integrating only the overlap with the absorber redshift window provides a conservative estimate of association probability and may down weight truly correlated galaxies. Hence this model should be regarded as a lower bound on the large-scale contribution, with the observed coincidence probabilities implying either a higher abundance of faint galaxy halos or broader redshift correlation than captured by our strict Gaussian weighting.


While the photometric-galaxy approach provides a fully empirical estimate of the clustering contribution, it is limited by photometric-redshift uncertainties and the finite depth of the imaging data. To verify that the inferred coincidence probabilities are not driven by these observational systematics, we complement this analysis with an independent theoretical estimate based on the galaxy two-point correlation function. Consider a quasar sightline (Q1) with a detected \mgii\ absorption system. Using $W_{2796}-b$ relation from Equation~\ref{eq:w-d}, the projected impact parameter between the absorber host galaxy and the Q1 sightline is inferred to be $b_1$. 
We then construct a cylindrical volume centered on the absorber, with a radius $r_{\mathrm{gal}}$ and a line-of-sight extent corresponding to a velocity interval of $\pm1500~\mathrm{km\,s^{-1}}$. 
The radius $r_{\mathrm{gal}}$ taken from Equation \ref{eq:w-d}, represents the radius associated with the adopted equivalent-width threshold, taking values of $r_{\mathrm{gal}} = 28.42$~kpc for $\mathrm{EW} > 1~\text{\AA}$ and $r_{\mathrm{gal}} = 40.10$~kpc for $\mathrm{EW} > 0.6~\text{\AA}$. A schematic diagram of this is shown in Figure \ref{fig:clustering_coincidecne_probability_diagrams} bottom panel.
The clustering-induced coincidence probability is then defined as the expected number of galaxies within this cylinder and is given by:
\begin{equation}
    P_{c} = \sum_{i=0}^{N} 
    \pi r_{gal}^{2}\, n_{0}\,\bigl[1+\xi(r_{i})\bigr]\, \Delta r_{\parallel} ,
    \label{eq:clustring_prob}
\end{equation}
where $r$ is $\sqrt{b_{2}^2 + r_{\parallel}^2}$, $\Delta r_{\parallel}$ the path–length element associated with the $i$th bin, $n_o$ is the average galaxy number density and $\xi(r)$ is the two point correlation function which is defined as:
\begin{equation}
    \xi(r) = \left( \frac{r}{r_{0}} \right)^{-\gamma},
\end{equation}
where $r$ is the distance of the point in space from the galaxy, $r_{0}$ is the comoving correlation length, and $\gamma$ is the slope of the correlation function.
We adopt the clustering parameters $r_{0}$ and $\gamma$ from measurements of galaxy clustering at $z \sim 1$ reported by \citet{Coil2004ApJ...609..525C}, based on the DEEP2 Galaxy Redshift Survey.
The DEEP2 sample is $R$-band selected with a limiting magnitude of $R_{\mathrm{AB}} < 24.1$, corresponding approximately to an absolute magnitude limit of $M_{B} \sim -20$ to $-20.5$ at $z \sim 1$, depending on galaxy spectral type.
The adopted values of $r_{0}$ and $\gamma$ therefore characterize the clustering of galaxies with luminosities near $L^{\ast}$ at these redshifts.
To calculate the galaxy number density relevant for Mg\,\textsc{ii} absorption, we derive $n_{0}$ from the rest--frame $B$--band Schechter luminosity function measured by \citet{Faber2007ApJ...665..265F}.
We adopt the parameters corresponding to the $z \simeq 1$ bin, with $\phi_{\ast} = 3.304 \times 10^{-3}~\mathrm{Mpc^{-3}}$, $M_{\ast} = -21.36$, and a fixed faint--end slope $\alpha = -1.30$, and propagate the quoted uncertainties in $\phi_{\ast}$ and $M_{\ast}$.
The Schechter function is integrated down to luminosity thresholds corresponding to three characteristic galaxy populations, namely $L^{\ast}$, $0.5L^{\ast}$, and $0.1L^{\ast}$, yielding comoving number densities appropriate for each luminosity bin.
These densities are subsequently converted to proper units at $z=1$ using
\begin{equation}
    n_{\mathrm{proper}} = n_{\mathrm{comoving}} (1+z)^3.
\end{equation}
The resulting clustering--based coincidence probabilities as a function of projected separation are shown in Figure~\ref{fig:clustering_curves}.
Among the three luminosity thresholds considered, the model assuming a characteristic luminosity of $0.1L^{\ast}$ provides the best overall agreement with the observed coincidence probability as a function of separation.
The models based on brighter galaxy populations ($L^{\ast}$ and $0.5L^{\ast}$) systematically under predict the observed signal at most separations.
We note that the coincidence probabilities inferred from photometric galaxy counting are consistently lower than the clustering--based predictions, which may reflect residual incompleteness in the photometric sample or uncertainties in the adopted covering fraction correction.
It is well established that the comoving correlation length evolves with
redshift as a consequence of the growth of large-scale structure and the
changing bias of the galaxy population. In general, clustering measurements
show weaker intrinsic matter clustering at higher redshift, while the
observed galaxy clustering reflects the combined effects of structure growth
and galaxy bias \citep[e.g.,][]{Coil2004ApJ...609..525C,Zehavi2011ApJ...736...59Z,Wang2021SCPMA..6489811W}, such that typical values of $r_{0}$ at $z \sim 1$ are smaller than those measured in the local Universe.
A smaller correlation length reduces the amplitude of the two--point correlation function, $\xi(r)$, thereby lowering the expected excess number of galaxies within the cylindrical search volume.
At the same time, when expressed in proper units, the galaxy number density increases with redshift according to $n_{\mathrm{proper}} = n_{\mathrm{comoving}}(1+z)^3$, which partially compensates for the reduced clustering amplitude.
In this work, we compute the clustering--based coincidence probability at $z \simeq 1$, corresponding to the median redshift of the Mg\,\textsc{ii} absorber sample, ensuring that the adopted clustering parameters and number densities are appropriate for the redshift range probed by our data.

\subsection{Stacking}

Next we  investigate coincident Mg\,\textsc{ii} absorption using stacking method. For this, we stack the spectrum of the second sightline for systems in which the first sightline contains a detected Mg~{\sc ii} absorber. Two sub-samples from the sample of 25764 pairs (see Table \ref{tab:samdet} line 3) are defined based on the rest-frame equivalent width measured in the first sightline, with thresholds of $W_{2796} > 0.6$\,\AA\ (22557 pairs) and $W_{2796} > 1.0$\,\AA~(16684 pairs). In addition, we perform the stacking analysis on the full sample without imposing any equivalent-width threshold in the first sightline.
For each system, the normalized spectrum of the opposite sightline is shifted to the absorber rest frame using the corresponding absorption redshift, and a $\pm50$\,\AA\ window centred on the Mg\,\textsc{ii} doublet is extracted. The continuum is again normalized by dividing by the median flux within this window. To estimate the noise level for each individual spectrum, we compute the root--mean--square (RMS) scatter of the flux residuals within this window, applying a $3\sigma$ clipping to remove outliers.
All spectra are interpolated onto a common rest--frame wavelength grid spanning 2750--2850\,\AA. The spectra are then grouped into projected separation (D) bins of width 200\,kpc and combined using an inverse--variance weighted mean, where the weights of pixels are defined by the inverse square of RMS noise of individual spectrum. This procedure yields a stacked flux profile for each separation bin. The statistical uncertainty on the stacked spectrum is estimated from the inverse square root of the summed weights, while an empirical error spectrum is also computed from the RMS scatter among the contributing spectra divided by the effective number of sightlines contributing to each wavelength bin. 

\begin{figure}
    \centering
    \includegraphics[width=\linewidth]{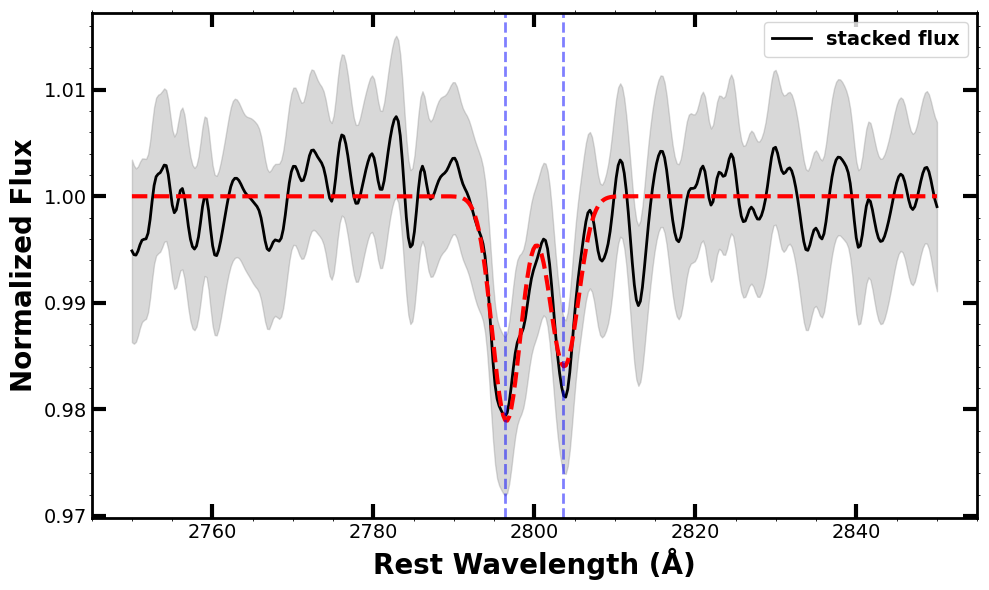}
    \includegraphics[width=\linewidth]{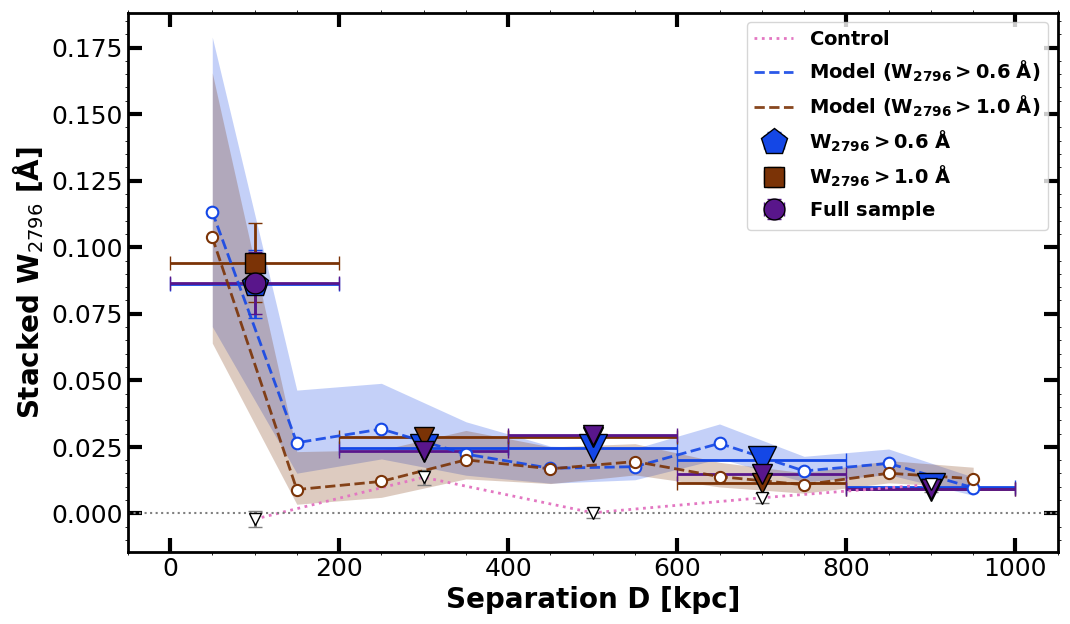}
    \caption{Top panel: Example stacked spectrum in the Mg\,\textsc{ii} $\lambda\lambda2796,2803$ region for pairs with projected separations $D = 0$--200~kpc. Bottom panel: Stacked Mg \textsc{ii} $\lambda2796$ equivalent width measured in the second sightline of pairs as a function of projected separation. Symbols show the observed stacked equivalent widths for systems in which the first sightline contains an absorber with $W_{2796} > 0.6$~\AA\ (blue, diamond) and $W_{2796} > 1.0$~\AA\ (brown, square), while purple symbols (filled circle) correspond to the full sample. Measurements at $D \le 200$~kpc represent statistically significant detections, whereas points at larger separations are shown as upper limits. Horizontal error bars indicate the projected-separation bin width (200 kpc) used in the stacking analysis. The dotted curve shows the result from a control sample constructed by stacking offset wavelength regions. Dashed curves show the corresponding model predictions derived from the observed coincidence probabilities (e.g., see Eq.~\ref{eq:stackew}).
}
    \label{fig:stack_plot}
\end{figure}

Each stacked spectrum is fitted with a two component Gaussian absorption model representing the Mg\,\textsc{ii} $\lambda\lambda2796,2803$ doublet, allowing for a common velocity width and a small global wavelength shift. The rest-frame equivalent widths are measured by integrating the absorption flux profile within $\pm3\sigma$ of the fitted line centres. This procedure yields the stacked Mg\,\textsc{ii} absorption strength in the second sightline as a function of projected separation (Figure~\ref{fig:stack_plot}). For projected separations $D < 200$\,kpc, we measure a stacked equivalent width of $W_{2796} = 0.0866 \pm 0.0116$\,\AA\ for the full sample, $W_{2796} = 0.0942 \pm 0.0148$\,\AA\ for the threshold $W_{2796} > 1.0$\,\AA\ subsample, and $W_{2796} = 0.0862 \pm 0.0126$\,\AA\ for the threshold $W_{2796} > 0.6$\,\AA\ subsample.
We find that the stacked equivalent width declines rapidly with increasing projected separation and becomes comparable to the amplitude of residual continuum fluctuations beyond $D \gtrsim 200$\,kpc. Visual inspection of the stacked spectra in these bins shows no statistically significant Mg\,\textsc{ii} doublet relative to the local spectral fluctuations arising from noise and imperfect continuum normalization. We therefore treat measurements at $D > 200$\,kpc as non--detections and present them as upper limits in our analysis.

As an additional check, we perform an identical stacking analysis on a control sample in which the same fitting procedure is applied to wavelength regions offset by 25~\AA~ from the Mg\,\textsc{ii} doublet. The resulting equivalent widths are consistent with zero at all separations, confirming that the weak signals observed at small projected separations are not produced by systematic effects in the stacking or fitting procedure.
We further repeat the stacking analysis by dividing the full sample into two redshift subsamples at the median absorber redshift, $z=1$. For projected separations $D<200$\,kpc, we measure a stacked equivalent width of $W_{2796} = 0.1151 \pm 0.0182$\,\AA\ for the high--redshift subsample ($z \ge 1$), compared to $W_{2796} = 0.0616 \pm 0.0142$\,\AA\ for the low--redshift subsample ($z < 1$). This indicates an increase in the stacked Mg\,\textsc{ii} absorption strength with redshift at small projected separations.

We model the stacked equivalent width at a given projected separation $D$ is as:
\begin{equation}
\langle W_{\mathrm{stack}}(D) \rangle
=
P(D)\int_{W_{\min}}^{W_{\max}} W\,p(W)\,dW ,
\label{eq:stackew}
\end{equation}
where $P(D)$ is the coincidence probability measured observationally (Figure~\ref{fig:coincidence probability}), and $p(W)$ is probability per unit equivalent width as defined in Equation~\ref{eq:ew_prob}. The $W_{min}$ and $W_{max}$ range corresponds to 0.6-4 ~\AA~ and 1-4~\AA~ for threshold $W_{2796}$> 0.6~\AA~ and $W_{2796}$>1~\AA~ sample respectively. Here, the dependence on projected separation enters entirely through the observed coincidence probability. Using the empirically measured $P(D)$ for the two equivalent-width thresholds ($W_{2796} > 0.6$~\AA\ and $W_{2796} > 1.0$~\AA), we compute model predictions for the stacked Mg\,\textsc{ii} absorption as a function of projected separation.
The resulting model predictions are shown in Figure~\ref{fig:stack_plot} and are compared directly with the measured stacked Mg\,\textsc{ii} equivalent widths. For projected separations $D \lesssim 200$~kpc, the model predicts a stacked equivalent width of $\langle W_{\mathrm{stack}} \rangle = 0.096^{+0.056}_{-0.036}$~\AA~ for the $W_{2796} > 0.6$~\AA~ threshold and $\langle W_{\mathrm{stack}} \rangle = 0.091^{+0.054}_{-0.034}$~\AA~ for the $W_{2796} > 1.0$~\AA~ threshold, in excellent agreement with the observed stacked measurements. The model successfully reproduces both the amplitude and the separation dependence of the stacked absorption signal, with the enhanced equivalent width at small separations arising naturally from the elevated coincidence probability in this regime. At larger separations, the flattening of the stacked equivalent width reflects the corresponding decline and plateau in $P(D)$. This close agreement indicates that the observed stacked Mg\,\textsc{ii} absorption is primarily governed by the probability of intersecting an absorber at a given projected separation.

\section{Discussion and Conclusions}
\label{sec:discussion}
The coincidence probability curve derived from our pair sample exhibits a
clear two–regime structure. At separations below $\sim$100\,kpc, the coincidence
probability rises sharply,
reaching values of $\sim$5--8\% (see Figure \ref{fig:coincidence probability}).   
This steep increase indicates that sightlines within $\lesssim$100\,kpc have a
significantly enhanced likelihood of intersecting the same
Mg\,\textsc{ii}-absorbing structure, consistent with both sightlines probing the
same dark-matter halo or an extended region of correlated cool gas within the
CGM.  
Beyond $\sim$100$-$200\,kpc, the coincidence probability declines rapidly and
settles into a low plateau of $\sim$1--2\% that persists out to nearly 1\,Mpc.

To interpret these trends, we compared the observed curve to simple physical
models.  
A single-halo model naturally reproduces the strong enhancement below
$\sim$100\,kpc, indicating that the high coincidence probability in this regime
originates from gas associated with a common halo and its extended CGM.  
At larger separations, where the coincidence probability remains nonzero but at
the $\sim$1--2\% level, the single-halo model cannot account for the
observed behaviour.  
We show that this low plateau is well explained by galaxy clustering: both
direct photometric galaxy counts and the two-point correlation function predict
a small but finite chance that two widely separated sightlines intersect
independent Mg\,\textsc{ii} absorbers hosted by galaxies correlated in
large-scale structure.  
Thus, the full coincidence curve can be understood as the sum of a halo-driven
component at small separations and a clustering-driven component at large
separations.

Our measured coincidence probabilities are lower than those reported by \citet{Tytler2009MNRAS.392.1539T}, who found coincidence fractions of $\gtrsim50\%$ at $D<100$~kpc that decline to $\sim0.8\%$ at $D\sim1$–$2$~Mpc. Despite this difference in normalization, the overall trend with transverse separation is qualitatively similar in both studies, with the coincidence probability decreasing rapidly with increasing separation and approaching very small values on Mpc scales. The higher coincidence fractions reported by \citet{Tytler2009MNRAS.392.1539T} likely arise from differences in sample selection and methodology, as their analysis focused primarily on higher-redshift ($z\sim2$) absorbers and included multiple metal species, particularly \ion{C}{IV}, which traces a more extended and highly ionized gas phase. In addition, their coincidence definition allowed matches between different ions (e.g., \ion{C}{IV}–\ion{Mg}{II}), whereas our analysis requires a strict \ion{Mg}{II}–\ion{Mg}{II} coincidence with uniform equivalent-width thresholds.

A recent study by \citet{Cortez2026arXiv260117127C} analyzed the clustering of \ion{C}{IV} absorbers using high-resolution spectra of quasar pairs and lensed quasars spanning transverse separations from sub-kiloparsec to megaparsec scales. They reported evidence for multiple coherence scales in the \ion{C}{IV} correlation function, with a small-scale break at $r_2 \approx 4.7^{+1.6}_{-1.2}$ kpc interpreted as the characteristic size of individual \ion{C}{IV}-bearing clouds, and a larger-scale break at $r_1 \approx 654$ kpc associated with the extent of enriched regions around galaxies. Between these scales the correlation function flattens, while it declines again at larger separations. In contrast, our Mg~\textsc{ii} coincidence probabilities remain approximately constant at large separations ($\gtrsim 200$ kpc), consistent with expectations from galaxy clustering. Part of this difference arises from the range of scales probed by the two studies. The small-scale break reported for \ion{C}{IV} occurs at a few kiloparsecs, which is below the minimum transverse separation accessible in our quasar-pair sample and therefore cannot be directly tested here. The larger-scale break in the \ion{C}{IV} correlation function occurs near $\sim650$ kpc, reflecting the more extended distribution of highly ionized gas. In our Mg~\textsc{ii} measurements the transition occurs at smaller scales ($\sim100$--200 kpc), marking the point where coincidences transition from being dominated by gas within a common halo to those produced by galaxy clustering. This difference is likely related to the distinct physical phases traced by the two ions. \ion{C}{IV} traces a warmer and more highly ionized gas phase that is expected to extend farther into the CGM and surrounding intergalactic environment, whereas Mg~\textsc{ii} predominantly traces denser, cooler gas confined to inner halo regions. As a result, \ion{C}{IV} correlations may remain sensitive to the structure of enriched regions on larger scales, while Mg~\textsc{ii} coincidences at similar separations are dominated primarily by the clustering of host galaxies rather than by coherent gas structures.

Our single-halo model predicts a high coincidence probability, reaching
$\sim80\%$ at separations $D\lesssim20$~kpc, reflecting the expected
coherence of Mg~\textsc{ii}-absorbing gas within the inner circumgalactic
medium. Although our quasar-pair sample does not probe such small
separations, independent studies of strongly lensed quasars provide
important constraints on the spatial structure of Mg~\textsc{ii}-bearing
gas on kiloparsec scales.
Observations of multi-image gravitational lenses frequently detect
Mg~\textsc{ii} absorption along multiple closely spaced sightlines,
indicating a high covering fraction and substantial coherence at small
separations \citep[e.g.,][]{Chen2014MNRAS.438.1435C,Zahedy2016MNRAS.458.2423Z,
Augustin2021MNRAS.505.6195A,Okoshi2021AJ....162..175O}. However, these
absorbers also exhibit significant sightline-to-sightline variations.
For example, \citet{Rogerson2012MNRAS.421..971R} showed that the observed
scatter in Mg~\textsc{ii} equivalent widths requires a small coherence
length ($\ell_c \simeq 0.5\,h^{-1}\,\mathrm{kpc}$), while other studies
report variations of $\lesssim40\%$ across separations of
$\sim8$–$22$~kpc \citep{Rubin2018ApJ...859..146R}.
Complementary constraints from gravitational-arc spectroscopy further
support this picture, revealing Mg~\textsc{ii} absorption across many
closely spaced sightlines and implying coherence on kiloparsec scales
\citep{Lopez2018Natur.554..493L,Afruni2023A&A...680A.112A,Shaban2025}.
These observational constraints are broadly consistent with
recent theoretical expectations for the structure of the cool CGM.
High-resolution cosmological simulations such as TNG50 predict that the
Mg~\textsc{ii}-bearing circumgalactic medium is composed of numerous
compact cloudlets with characteristic sizes of order $\sim$kpc or
smaller \citep{Nelson2020MNRAS.498.2391N}, while zoom-in simulations
likewise show that Mg~\textsc{ii} absorption can arise from dense
small-scale structures embedded within halo gas
\citep{Ramesh2024MNRAS.528.3320R}. Spectral modeling studies also infer
a fragmented multiphase CGM in which multiple discrete clouds contribute
to the absorption along a single sightline
\citep{Sameer2024MNRAS.530.3827S}.
Taken together, these observations suggest that Mg~\textsc{ii} absorption
arises from numerous small-scale structures within galaxy halos,
producing a combination of high covering fraction and significant
small-scale variability. This picture is broadly consistent with the
high coincidence probabilities predicted by our single-halo model at
small impact parameters.

It is important to note, however, that our single-halo model is constructed using the observed $W$–$D$ relation and therefore describes the average radial behavior of Mg~\textsc{ii} absorption, implicitly assuming a smooth distribution of absorbing gas at fixed impact parameter. As such, the model does not explicitly incorporate small-scale spatial inhomogeneities or a clumpy cloudlet structure in the CGM. The strong sightline-to-sightline variations observed in lensed quasar systems instead suggest that Mg~\textsc{ii}-bearing gas is highly fragmented on kiloparsec and sub-kiloparsec scales. In this context, the high coincidence probabilities predicted by our model at small separations should be interpreted as reflecting the ensemble-averaged covering fraction of such cloudlets, rather than a spatially coherent medium.

Our stacking analysis provides independent support for this picture.  
For sightlines lacking individual Mg\,\textsc{ii} detections, stacked spectra
show a significant excess in mean equivalent width on the side where the paired
sightline does not exhibits absorption.  
This enhancement extends to separations of $\sim$100--200\,kpc, suggesting that
Mg\,\textsc{ii} absorption is correlated across these scales even when the
individual features fall below the detection threshold.  
The stacking therefore constrains the characteristic length of the
Mg\,\textsc{ii}-bearing CGM to be of order $\sim$100--200\,kpc, consistent with
the interpretation of the coincidence curve. Using the observed coincidence probability, we show that the measured stacked equivalent widths are naturally reproduced by a simple physically motivated model. Our stacked measurements at small projected separations are broadly consistent
with independent DESI-based stacking analyses of Mg\,\textsc{ii} absorption
around galaxies at comparable redshifts \citep{chen2025circumgalacticmediumtracedmg}.
Restricting our sample to projected separations $D < 200$~kpc, the median
separation is $D_{\rm med} \simeq 140$~kpc, at which we measure a stacked
equivalent width of $W_{2796} \sim 0.1$~\AA.
At comparable projected separations ($\sim$140~kpc), the DESI-based stacking
analysis of \citet{chen2025circumgalacticmediumtracedmg} reports similar
Mg\,\textsc{ii} equivalent widths of order $W_{2796} \sim 0.1$~\AA\ for galaxies
at $z \sim 1$.
This agreement indicates that, despite differences in sample selection and
methodology, both studies recover a consistent amplitude and radial decline of
Mg\,\textsc{ii} absorption within the inner $\sim$200~kpc of the CGM.

Together, these results indicate that Mg\,\textsc{ii} absorption exhibits strong sightline correlations within the inner $\sim$100\,kpc of
galaxy halos, while at larger separations the correlations are dominated by
galaxy clustering rather than by gas confined to a single halo.  
Our measurements therefore identify the transition scale between the halo regime
and the large-scale-structure regime, bridging the gap between sub-kpc
coherence constraints from lensed quasars and Mpc-scale absorber clustering
studies.  
Future work will require substantially larger samples of quasar pairs at smaller transverse separations in order to more densely sample the rising portion of the coincidence and stacking curves. Such observations will enable a more precise characterization of the spatial coherence, characteristic scales, and internal structure of the Mg~\textsc{ii}-bearing cool CGM in the inner halo regime.

\section*{Acknowledgements}
The research of P.S. is supported by the University Grants Commission (UGC), Government of India, under the UGC-JRF scheme  (Ref. No.: 221610014755). H.C. and P.S. express their gratitude to the Inter-University Centre for Astronomy and Astrophysics (IUCAA) for their hospitality and the provision of High-Performance Computing (HPC) facilities under the IUCAA Associate Programme. Additionally, we acknowledge the assistance of AI tools, specifically OpenAI's ChatGPT, for aiding in writing and code development, and Grammarly for enhancing the text's clarity and correctness.
\section*{Data Availability}
The data used in this study are publicly available in the SDSS DR16 Data Release.


\bibliographystyle{mnras}
\bibliography{references} 








\bsp	
\label{lastpage}
\end{document}